\documentclass[aps,prl,preprintnumbers,showpacs,nofootinbib]{revtex4}

\usepackage{graphicx}
\usepackage{dcolumn}
\usepackage{bm}
\begin{document}
\newcommand{\newc}{\newcommand}
\newc{\ra}{\rightarrow}
\newc{\lra}{\leftrightarrow}
\newc{\beq}{\begin{equation}}
\newc{\eeq}{\end{equation}}
\newc{\barr}{\begin{eqnarray}}
\newc{\earr}{\end{eqnarray}}
\newcommand{\Od}{{\cal O}}
\newcommand{\lsim}   {\mathrel{\mathop{\kern 0pt \rlap
  {\raise.2ex\hbox{$<$}}}
  \lower.9ex\hbox{\kern-.190em $\sim$}}}
\newcommand{\gsim}   {\mathrel{\mathop{\kern 0pt \rlap
  {\raise.2ex\hbox{$>$}}}
  \lower.9ex\hbox{\kern-.190em $\sim$}}}

\title{Direct detection of supersymmetric dark
matter-
 Theoretical rates for transitions to excited states.}

\author{J.D. Vergados$^{a,c}$, P. Quentin$^{b}$ and D. Strottman$^{b}$}

\affiliation{$^a$ Theoretical Physics Division, T-6, LANL, P.O.
Box 1663,
Los Alamos, N.M. 87545.\\
$^b$ Theoretical Physics Division, T-DO, LANL, P.O. Box 1663,
Los Alamos, N.M. 87545.\\
$^c$Theoretical Physics Division, University of Ioannina, Gr 451
10, Ioannina, Greece.}
\vspace{0.5cm}
\begin{abstract}
The recent WMAP data have confirmed that exotic dark matter
together with the vacuum energy (cosmological constant) dominate
in the flat Universe. Supersymmetry provides a natural dark matter
candidate, the lightest supersymmetric particle (LSP).
 Thus the direct dark matter detection is central to
particle physics and cosmology.  Most of the research on this
issue has hitherto focused on the detection of the recoiling
nucleus. In this paper we study transitions to the excited states,
focusing on the first excited state at 50 keV of Iodine A=127. We
find that the transition rate to this excited state is $\preceq
10$ percent of the transition to the ground state. So, in
principle, the extra signature of the gamma ray following its
de-excitation can be exploited experimentally.
\end{abstract}

\date{\today}
\maketitle
\section{Introduction}
The combined MAXIMA-1 \cite{MAXIMA-1}, BOOMERANG \cite{BOOMERANG},
DASI \cite{DASI}, COBE/DMR Cosmic Microwave Background (CMB)
observations \cite{COBE}, the recent WMAP data \cite{SPERGEL} and
SDSS
 \cite{SDSS} imply that the
Universe is flat \cite{flat01} and
 and that most of the matter in
the Universe is dark, i.e. exotic.
  $$ \Omega_b=0.044\pm 0.04,
\Omega_m=0.27\pm 0.04,  \Omega_{\Lambda}=0.69\pm0.08$$
 for baryonic matter , cold dark matter and dark energy
respectively. An analysis of a combination of SDSS and WMAP data
yields \cite{SDSS} $\Omega_m\approx0.30\pm0.04(1\sigma)$. Crudely
speaking
$$\Omega_b\approx 0.05, \Omega _{CDM}\approx 0.30, \Omega_{\Lambda}\approx 0.65$$

Since the non exotic component cannot exceed $40\%$ of the CDM
~\cite {Benne}, there is room for the exotic WIMP's (Weakly
Interacting Massive Particles).
  In fact the DAMA experiment ~\cite {BERNA2} has claimed the observation of one signal in direct
detection of a WIMP, which with better statistics has subsequently
been interpreted as a modulation signal \cite{BERNA1}, although
these data are not consistent with other recent experiments, see
e.g. EDELWEISS \cite{EDELWEISS} and CDMS \cite{CDMS}.

 Supersymmetry naturally provides candidates for the dark matter constituents
\cite{Jung},\cite{GOODWIT}-\cite{ELLROSZ}.
 In the most favored scenario of supersymmetry the
LSP can be simply described as a Majorana fermion, a linear
combination of the neutral components of the gauginos and
higgsinos \cite{Jung},\cite{GOODWIT}-\cite{ref2}. In most
calculations the neutralino is assumed to be primarily a gaugino,
usually a bino. Models which predict a substantial fraction of
higgsino lead to a relatively large spin induced cross section due
to the Z-exchange. Such models tend to violate the LSP relic
abundance constraint and are not favored.  Some claims have
recently been made, however, to the effect that the WMAP relic
abundance constraint can be satisfied in the hyperbolic branch of
the allowed SUSY parameter space, even though the neutralino is
then primarily a higgsino \cite{CCN03}. We will not elaborate
further on this point, but we will  take the optimistic view that
the detection rates due to the spin may be large enough to be
exploited by the experiments, see, e.g., \cite{CCN03}
\cite{CHATTO}, \cite{WELLS}, \cite{JDV03} . Such a view is further
encouraged by the fact that, unlike the scalar interaction, the
axial current allows one to populate excited nuclear states,
provided that their energies are sufficiently low so that they are
accessible by the low energy LSP, a prospect proposed long time
ago by Ejiri and collaborators \cite{EFO93}. As a matter of fact
the average kinetic energy of the LSP is:
 \beq <T> \approx40~keV
\frac{m_{\chi}}{100~GeV}
 \label{kinen}
 \eeq
 So for sufficiently heavy LSP the
average energy may exceed the excitation energy, e.g. of about
$50~keV$ for the excited state of $^{127}I$. In other words one
can explore the high velocity window, up to the escape velocity of
$\upsilon_{esc}\approx 570~km/s$. From a Nuclear Physics point of
view this transition is not expected  to be suppressed, since it
is of the type $(5/2)^{+} \rightarrow (7/2)^{+}$, i.e.
Gamow-Teller like.
\section{The LSP-Nucleus Cross Section}
 The LSP-nucleus  differential
cross section with respect to the energy transfer $Q$ for a given
LSP velocity $\upsilon$ due to the spin can be cast in the form
\begin{equation}
d\sigma (u,\upsilon)= \frac{du}{2 (\mu _r b\upsilon )^2}
\bar{\Sigma} _{spin} F_{11}(u) \label{2.9}
\end{equation}
where $F_{11}$ is the spin response of the isovector channel
\cite{DIVA00}. We have used a dimensionless variable $u$,
proportional to $Q$, which has been  found convenient for handling
the nuclear form factor \cite{KVprd} F(u), namely:
 \beq
u=\frac{Q}{Q_0}~~,~~Q_{0}=\frac{1}{Am_N b^2}\approx 40 \times
A^{-4/3}~MeV.
 \label{Qu}
 \eeq
  $\mu_r$
is the reduced LSP-nucleus mass and $b$ is the (harmonic
oscillator) nuclear size parameter.

The quantity $\bar{\Sigma} _{spin}$ is given by:
\begin{equation}
\bar{\Sigma} _{spin}  =  (\frac{\mu_r}{\mu_r(p)})^2
                           \sigma^{spin}_{p,\chi^0}~\zeta_{spin},
\zeta_{spin}= \frac{1}{3(1+\frac{f^0_A}{f^1_A})^2}S(u)
\label{2.10a}
\end{equation}
$\mu_r(p)\approx m_p$ is the LSP-nucleon reduced mass and
$\sigma^{spin}_{p,\chi^0}$ is the proton
 cross-section associated with the spin and $S(u)$ \cite{DIVA00} is given by:
\begin{equation}
S(u)=[(\frac{f^0_A}{f^1_A} \Omega_0(0))^2
\frac{F_{00}(u)}{F_{11}(u)}
  +  2\frac{f^0_A}{ f^1_A} \Omega_0(0) \Omega_1(0)
\frac{F_{01}(u)}{F_{11}(u)}+  \Omega_1(0))^2  \, ]
\label{spin}
\end{equation}
 The  overall normalization of the axial current components
 is not important in the present
 discussion. We mention, however, that they have been  normalized so that
 $\sigma^{spin}_{p,\chi^0}=3(f_A^0+F_A^1)^2 \sigma_0$,
  $\sigma^{spin}_{n,\chi^0}=3(f_A^0-F_A^1)^2 \sigma_0$, $\sigma_0=\frac{1}{2 \pi}(G_F
  m_p)^2$,
  for the proton and neutron spin cross sections
  respectively.

Notice that $S(u)$, being dependent on ratios, is expected to be
less dependent on the energy transfer and the scale of
supersymmetry. As a matter of fact it has been found \cite{DIVA00}
that the spin response functions $F_{ij}$, the "spin form factors"
normalized to unity at $u=0$, have similar dependence on the
energy transfer. In other words $S(u)$ is essentially independent
of $u$.

 A number of nuclear spin matrix elements for a variety of targets have become
available \cite{DIVA00}-\cite{SUHONEN03}. Some  static spin matrix elements
 \cite{DIVA00}, \cite{Ress}, \cite{KVprd}
for some nuclei of interest are given in
table \ref{table.spin}.
The spin matrix elements appearing in the table are defined as:
 \beq
 \Omega_0=\Sigma_p+\Sigma_n~~, \Omega_1=\Sigma_p-\Sigma_n~~,~~
 \Sigma_k=\frac{1}{\sqrt{2J_i+1}}<J_f||2~{\bf S}_k||J_i>,~~k=p,n~,
  \label{sigmame}
  \eeq
 where by double bar we indicate the reduced matrix element as defined in
standard textbook as e.g the one by Edmonds. The  $<S_k>~~,k=p,n$,
are defined in the literature, e.g. Ressel
 and Dean \cite{Ress}.

 In order to proceed, in addition to the static spin matrix
 elements and the spin response functions $F_{00},F_{01},F_{11}$
 ("spin form factors"), normalized to unity at zero momentum transfer, one
 must know the isoscalar and the isovector axial currents.
  In the allowed supersymmetric parameter space one obtains the isoscalar and
 the isovector
axial current components at the quark level. Going from the quark
to the nucleon level is straightforward for the isovector current,
but it not trivial in the case of the isoscalar current, since
 the naive quark model fails badly. Most of the proton spin is not due to the quark spins
 (proton spin crisis-EMC effect). Thus one finds:
  \beq
 f^0_{A}(q) \rightarrow f^0_{A}= g^0_A~f^0_A(q)~~,
 f^1_{A}(q) \rightarrow f^1_{A}= g^0_A~f^1_A(q)~~,
g^0_A\approx0.1~~,~~g^1_A=1.23
 \label{fA}
  \eeq
  It thus appears that, barring very special circumstances, whereby
  the isoscalar contribution at the quark level, $f_A^0(q)$, is much larger than
  the isovector, $f_A^1(q)$, the
 isoscalar contribution can be neglected.

\section{The structure of $^{127}I$}
This nucleus has a ground state $5/2^+$ and a first excited state
a $7/2^+$ at $57.6 keV$. As it has already been mentioned  it is a
popular target for dark matter detection.  As a result the
structure of its ground state has been studied theoretically by a
lot of groups. Among them we mention again the work
 of Ressel and Dean \cite{Ress}, the work of Engel, Pittel and Vogel
\cite{PITTEL94}, Engel and Vogel \cite{ENGEL89}, Iachello, Krauss
and Maiano \cite {IACHELLO91}, Nikolaev and Klapdor-Kleingrothaus
\cite{NIKOLAEV93} and more recently by Suhonen and collaborators
\cite{SUHONEN03}. In all these calculations it appears that the
spin matrix element is dominated by its proton component, which in
our notation implies that the isoscalar and the isovector
components are the same. In these calculations there appears to be
a spread in the spin matrix elements ranging from $0.07$ up to
$0.354$, in the notation of Ressel and Dean \cite{Ress}. This, of
course, implies discrepancies of about a factor of 25 in the event
rates.

 In the present work we are primarily interested in the spin matrix element
connecting the ground state to the first excited state. As we have
mentioned, however, it is advantageous to compute the branching
ratio. In addition to factoring out most of the uncertainties
connected with the SUSY parameters and the structure of the
nucleon, we expect the ratio of the two spin matrix elements to be
more reliable than their absolute values. Thus this ratio is going
to be the most important factor in designing the experiments to
detect the rate to the excited state.

 In order to have for $Z=53$ single particle orbits suitable for the above states,
 close to the Fermi surface \cite{NILSSON}, namely  $[402,5/2]$ and $[404,7/2]$ proton orbits,
 one should consider oblate deformation \cite{BOHRMO}. The value of the deformation, however, is not
 very well defined.  To this end one should explore the spectrum of this system and combine
 information obtained from it with data on the neighboring systems $Te$ and $Xe$.
 On notices that on the above  band heads $\Delta J=2$ rotation aligned coupling bands are built
 \cite{SHROY}. From the spectra one may also conclude that the $7/2^+$ data most likely
 correspond to a deformed band. The $5/2$ data correspond to a highly perturbed band
 or to a spherical vibrator. Other people have considered triaxial shapes, see ref. 9 of Shroy
 {\it et al} \cite{SHROY}. So we will take it as an educated guess that a description in
 terms of a rotor plus a particle with relatively large oblate deformation provides
  a reasonable description of this nucleus.

  Given the above framework one may eventually have to do a
  self-consistent calculation, which will yield both the single
  particle states and the equilibrium deformation. At this exploratory stage, however, we were merely
  interested in testing the rather crude collective model contentions discussed above. To that effect
  we found it appropriate to take as simple, and easy to read, as possible the single particle substrate
  of our description. Thus to estimate the
  spin matrix elements, in the present simple spectroscopic study, we have proceeded in the
   following oversimplified fashion:
\begin{itemize}
\item Make a big leap in history to Nilsson`s thesis
\cite{NILSSON}.
 \item Compute the spin proton matrix elements involving the above bare single
 particle states for three choices of oblate deformations.
 \item include the effect of the Coriolis coupling for the $7/2^+$.
 \end{itemize}
 The relevant deformation parameter $\eta$ is defined \cite{NILSSON} as
 \begin{equation}
 \eta \approx \frac{\delta \left (1-(4/3)\delta^2 \right )}{\chi}
 \label{deform1}
 \end{equation}
 where $\delta$ is the usual Nilsson deformation parameter and
 $\chi \approx 0.05$. The values of $\eta$ considered are
 $-6,-4,-2$. To first order the corresponding $\delta$ values are:
 $-0.3,-0.2,-0.1$. The corresponding states in the standard notation
 $N\ell\Lambda\Sigma$ are given by:
  \beq
 |5/2>=\frac{ a|442+>+b|422+>+c|443->}{\sqrt{a^+b^2+c^2}}
  \label {2,5}
  \eeq
  \beq
 |7/2>=\frac{
  d|443+>+e|444->}{\sqrt{d^2+e^2}}
  \label{3.5}
  \eeq
 The numerical values for these coefficients in the above order of
 the $\eta `s$ are:
 $$\eta=-6 \rightarrow (a,b,c,d,e)=(1.000,0.552,-0.809,1.000,-1.192)$$
 $$\eta=-4 \rightarrow (a,b,c,d,e)=(1.000,0.558,-1.081,1.000,-1.662)$$
 $$\eta=-2 \rightarrow(a,b,c,d,e)=(1.000,0.517,-1.451,1.000,-2.219)$$
 Thus in the standard angular momentum notation $|J,M>$ one
obtains:
 \beq
 <5/2,5/2|s_z| 5/2,5/2>=0.166,0.033,-0.124
 \label{sz1}
 \eeq
 \beq
<7/2,7/2|s_z |7/2,7/2>=-0.087,-0.234,-0.331
 \label{sz2}
  \eeq
 \beq
<5/2,5/2|s_+| (7/2,5/2)>=0.171,-0.120,-0.073
 \label{sz3}
 \eeq
  The array of numbers in the above three equations
  corresponds to  $\eta=-6,-4,-2$ respectively.

One may also have to consider the effect of the Coriolis coupling,
see e.g. \cite{SHALIT}, between the $I=7/2$ and the $K=5/2,7/2$
bare single particle states with the Hamiltonian matrix as
follows:
 \beq
M_{ii}=E_i^{bare}+\frac{\hbar ^2}{2J} \left [I(I+1)-2K^2 \right]
 \label{cor1}
  \eeq

for the diagonal term. For the off-diagonal term one has:
 \beq
 V=\frac{\hbar^2}{J} \left [\frac{I(I+1)-K_{<}(K_{<}+1)}{2}
 \right]^{1/2}
 \left < K_{>}|j_+ |K_{<} \right >
  \label{cor2}
  \eeq
 where $j_{+}$ is given in units of $\hbar$ . One, of course,
 has to adopt a reasonable value for the moment of
 inertia $J$.
  For such a choice and $\eta=-6$ one finds:
   \beq
  M_{11}=387,M_{22}=145,
   V=207
   \label{mv}
   \eeq
   (in $KeV$).
  The resulting components of the eigenfunctions are:

   \hspace{2.0cm}$0.50,0.87$ and $-0.87,0.50$ for the $5/2$
  and $7/2$ respectively.

With the above wave functions   we find the ground state spin
matrix elements:
$$<S_p>=0.166~~,~~<S_n>=0$$,
 which are in good
agreement with Iachello {\it et al} \cite{IACHELLO91} and Nikolaev
and Klapdor-Kleingrothaus \cite{NIKOLAEV93}, but substantially
smaller than Ressel and Dean \cite{Ress}. In our notation this
leads to:
 \beq
\Omega_0^2= \Omega_1^2 =\Omega_0 \Omega_1=0.164
 \label{omegas0}
 \eeq
 This value is also smaller than the more recent result \cite{SUHONEN03}.
For the transition to the excited state we have obtained:
 \beq
\Omega_0^2= \Omega_1^2 =\Omega_0 \Omega_1=0.312
 \label{omegas1}
 \eeq
  Thus the  spin
matrix element, left to itself, seems to yield an enhancement of
about a factor of two for the transition to the excited state. In
our simple model this result is independent of the ratio of the
isoscalar to isovector amplitude.
\section{Expressions for the rates}
 In this section we will briefly outline the expressions yielding
  the event rates, both modulated and unmodulated,
 (for more details see our earlier work \cite{JDV03} and references
 therein).
 The differential non directional  rate can be written as
\begin{equation}
dR = \frac{\rho (0)}{m_{\chi}} \frac{m}{A m_N}
 d\sigma (u,\upsilon) | {\boldmath \upsilon}|
\label{2.18}
\end{equation}
where $d\sigma(u,\upsilon )$ was given above, $\rho(0)$ is the LSP
density in our vicinity, $m_{\chi}$ the LSP mass, $m$ is the mass
of the target and
$A$ is the nuclear mass number.\\
For a given velocity distribution f(\mbox{\boldmath
$\upsilon$}$^{\prime}$),
 with respect to the center of the galaxy,
one can find the  velocity distribution in the lab frame
$f(\mbox{\boldmath $\upsilon$},\mbox{\boldmath $\upsilon$}_E)$ by
writing

\hspace{2.0cm}\mbox{\boldmath $\upsilon$}$^{'}$=
          \mbox{\boldmath $\upsilon$}$ \, + \,$ \mbox{\boldmath $\upsilon$}$_E
 \,$ ,
\hspace{2.0cm}\mbox{\boldmath $\upsilon$}$_E$=\mbox{\boldmath
$\upsilon$}$_0$+
 \mbox{\boldmath $\upsilon$}$_1$

where \mbox{\boldmath $\upsilon_0=229~km/s$}$\,$  is the sun's
velocity (around the center of the galaxy) and \mbox{\boldmath
$\upsilon$}$_1 \,$ the  Earth's velocity
 (around the sun).
 The velocity of the earth is given by
\begin{equation}
\mbox{\boldmath $\upsilon$}_E  = \mbox{\boldmath $\upsilon$}_0
\hat{z} +
                                  \mbox{\boldmath $\upsilon$}_1
(\, sin{\alpha} \, {\bf \hat x} -cos {\alpha} \, cos{\gamma} \,
{\bf \hat y} + cos {\alpha} \, sin{\gamma} \, {\bf \hat z} \,)
\label{3.6}
\end{equation}
In the above formula $\hat{z}$ is in the direction of the sun`s
motion, $\hat{x}$ is in the radial direction out of the galaxy,
$\hat{y}$ is perpendicular in the plane of the galaxy
($\hat{y}=\hat{z} \times \hat{x}$) and $\gamma \approx \pi /6$ is
the inclination of the axis of the ecliptic
 with respect to the plane of the galaxy. $\alpha$ is the phase of the Earth
in its motion around the sun ($\alpha=0$ around June 2nd). The
time dependence of the event rate, modulation effect, will be
present for transitions to the excited states as well.

 Folding with the velocity distribution we get:
\begin{equation}
\Big<\frac{dR_{undir}}{du}\Big> = \Big<\frac{dR}{du}\Big> =
\frac{\rho (0)}{m_{\chi}} \frac{m}{Am_N} \sqrt{\langle
\upsilon^2\rangle } {\langle \frac{d\Sigma}{du}\rangle }
\label{3.11}
\end{equation}
where
\begin{equation}
\langle \frac{d\Sigma}{du}\rangle =\int
           \frac{   |{\boldmath \upsilon}|}
{\sqrt{ \langle \upsilon^2 \rangle}}
 f(\mbox{\boldmath $\upsilon$},\mbox{\boldmath $\upsilon$}_E)
                       \frac{d\sigma (u,\upsilon )}{du} d^3
 \mbox{\boldmath $\upsilon$}
\label{3.12a}
\end{equation}

In our calculation we have used the standard M-B distribution:
\begin{equation}
 f( \mbox{\boldmath $\upsilon$} ^{\prime})=\frac{1}{(\pi\upsilon_0^2)^{(3/2)}}
 Exp[-\frac{(\mbox{\boldmath
 $\upsilon$} ^{\prime})^2}{\upsilon_0^2}]
 \label{M-B1}
 \end{equation}
 In what follows we will express the LSP velocity in units
 of $\upsilon_0$ introducing the dimensionless quantity $y=\upsilon/ \upsilon _0$.
 Incorporating the relevant kinematics  and integrating
  the above expression, Eq. (\ref{3.12a}), from $u_{min}$ to $u_{max}$ we
 obtain the total rates as follows:
 \beq
 R_{gs}=\int_{u_{min}}^{u_{max}}\Big<\frac{dR_{gs}}{du}\Big>du
 \label{Rgs}
 \eeq
 with $u_{min}=Q_{min}/Q_0$, $Q_{min}$ imposed by the detector
 energy cutoff and $u_{max}=(y_{esc}/a)^2$ , with $a=[\mu_rb\sqrt{2}]^{-1}$,
 is imposed by the escape velocity ($y_{esc}=2.84$).
 \beq
 R_{exc}=\int_{u_{exc}}^{u_{max}}\Big<\frac{dR_{exc}}{du}(1-\frac{u^2_{exc}}{u^2})\Big>du
 \label{Rexc}
 \eeq
 where $u_{exc}=\frac{\mu_rE_{x}}{Am_NQ_0}$ and $E_{x}$ is the
 excitation anergy of the final nucleus. The upper limit is now given
 by $u_{max}=(y/a)^2-(E_x/Q_0)$
\section{Branching ratios}

In the present calculation, to spare the reader with details about
the SUSY allowed parameter space, we are not going to discuss the
absolute value of the event rates. We will, instead, estimate the
ratio of the rate
 to the excited state divided by that to the ground state
 (branching ratio of the rates) as a function of the LSP mass.
  This is of interest to the
 experimentalists. It can be cast in the form:
\begin{equation}
BRR =  \frac{S_{exc}(0)}{S_{gs}(0)}
    \frac{\Psi_{exc}(u_{exc},u_{umax})[1+h_{exc}(u_{exc},u_{max})~\cos{\alpha}]}
           {\Psi_{gs}(u_{min})[1+h(u_{min})~\cos{\alpha}]}
\label{rR}
\end{equation}
In the above expression $h$ and $h_{exc}$ are the modulation
amplitudes for the ground state and excited state respectively.
They correspond to the ratio of the amplitude associated with the
motion of the Earth divided by the one which is independent of the
motion of the Earth. The parameter $\alpha$ is the phase of the
Earth in its motion around the Sun ($\alpha=0$ on June 3nd). The
modulation amplitudes  can only be obtained numerically and  are
not going to be discussed
 in detail here (for more information see our earlier work \cite{JDV03}).

$S_{gs}(0)$ and $S_{exc}(0)$ are the static spin matrix elements
 evaluated
via Eq. \ref{spin}. This ratio of the static spin matrix elements
is essentially independent of supersymmetry, if the isoscalar
contribution is neglected due to the EMC effect. The functions
$\Psi$ are given as follows :
\begin{equation}
\Psi_{gs} (u_{min}) =\int _{u_{min}}^{(y/a)^2} \frac{S_{gs}
(u)}{S_{gs} (0)} F^{gs}_{11}(u)
             \big[ \psi (a \sqrt{u}) - \psi (y_{esc}) \big] ~du
\label{Psigs}
\end{equation}
\begin{equation}
\Psi_{exc} (u_{exc},u_{max}) =\int _{u_{exc}}^{u_{max}}
\frac{S_{exc} (u)}{S_{exc} (0)}
F^{exc}_{11}(u)(1-\frac{u^2_{exc}}{u^2})
             \big[ \psi (a \sqrt{u}(1+u_{exc}/u)) - \psi (y_{esc}) \big] ~du
\label{Psi1}
\end{equation}
The functions $\psi$ were evaluated numerically, with the correct
kinematics imposed on the neutralino velocity in the laboratory,
which is obtained by translating the simple relation $\upsilon^{'}
\le \upsilon_{esc}$, which holds with respect to the galactic
center. To a very good approximation, however, they can be given
by:
\begin{equation}
\psi (z)=\frac{1}{2}\big[ Erf(z+1)-Erf(z-1) \big] \label{psi}
\end{equation}
and with $Erf(x)$ the well known error function:
$$Erf(x)=\frac{2}{\sqrt{\pi}}\int_0 ^x e^{-t^2} ~dt.$$
 As we have mentioned the dependence of $S(u)$ on the energy
transfer is extremely mild and it can be neglected.

\section{Conclusions}
\begin{figure}
\begin{center}
\rotatebox{90}{\hspace{1.0cm} {\tiny BRR}$\rightarrow$}
\includegraphics[height=.10\textheight]{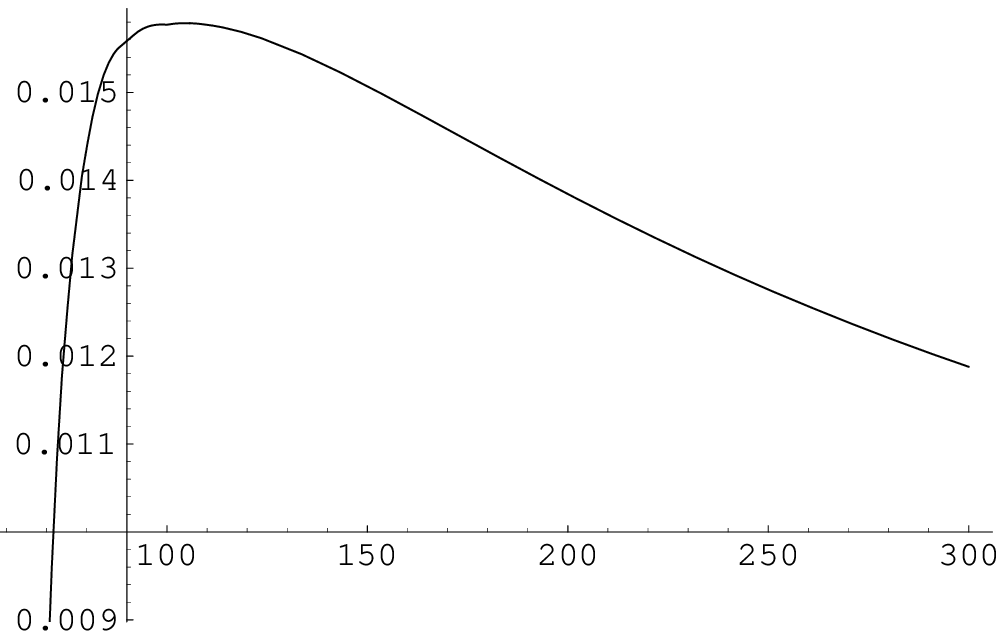}
 \hspace{0.0cm} $m_{LSP}\rightarrow$ ($GeV$)
 \rotatebox{90}{\hspace{1.0cm} {\tiny BRR}$\rightarrow$}
\includegraphics[height=.10\textheight]{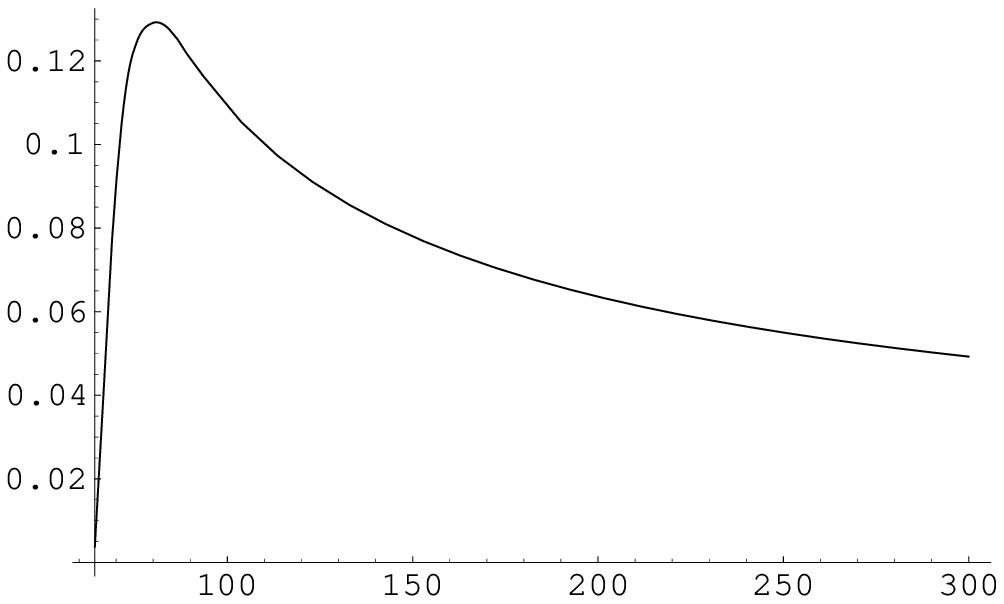}
 \hspace{0.0cm}$m_{LSP}\rightarrow$ ($GeV$)
 \caption{ The ratio of the rate to
the excited state divided by that of the ground state as a
function of the LSP mass (in GeV) for $^{127}I$. We assumed that
the static spin matrix element of the transition from the ground
to the excited state is a factor of 1.9 larger than that
involving the ground state, but  the spin response functions
$F_{11}(u)$ are
the same. On the left we show the results for $Q_{min}=0$ and on
the right for $Q_{min}=10~KeV$. \label{ratio} }.
\end{center}
\end{figure}

In the present paper we have calculated the branching ratio BBR
for the neutralino - $^{127}$I scattering event rate to the first
excited state. We found that  the ratio of the static spin matrix
element
 involving the excited state is $1.9$ times larger than that of the
 ground state. Assuming that the spin response functions $F_{11}(u)$
are
identical, we obtained  the ratio BRR (branching ratio of the rate
to the excited divide by that to the ground state), which is
exhibited in Fig. \ref{ratio}. In computing these branching ratios
we did not include the quenching factors affecting the nuclear
recoils, i.e. the detected event rates for transitions to the
ground state. If such factors are included, the branching ratios
will increase.

The difference in the branching ratios of   Fig. \ref{ratio} can
be understood as follows: Since the transitions the excited state
will be detected not by nuclear recoil, but via the EM decay of
the final state, they do not suffer from the threshold energy cut
off. This cut off suppresses  only the transitions to the ground
state, which are detected through nuclear recoils. Anyway the
branching ratio is affected. Thus this ratio is very small in the,
at present, unattainable experimentally case of $Q_{min}=0$. As we
have mentioned, however, even in this case, the branching ratio
will increase, if the quenching factor is taken into account. The
branching ratio is small even in the case of $Q_{min}=10$ keV for
LSP masses less than $50$ GeV. This is not unexpected, since such
a light neutralino cannot provide the energy needed to excite the
nucleus.
 For $Q_{min}=10$ this branching ratio attains a maximum value, around $10\%$, for
  LSP masses around  $100$ GeV
 and gets values around $6\%$ for larger masses. For larger $Q_{min}$
 the branching ratio will further increase.

 The  smallness of the
branching ratio, $\preceq 10\%$, should not discourage the
experiments, since the transition to the excited state offers
experimental advantages. This is especially true, if the
traditional experiments cannot reach $Q_{min}\preceq 10$ keV.

\begin{figure}
\begin{center}
\rotatebox{90}{\hspace{1.0cm} $h\rightarrow$}
\includegraphics[height=0.1\textheight]{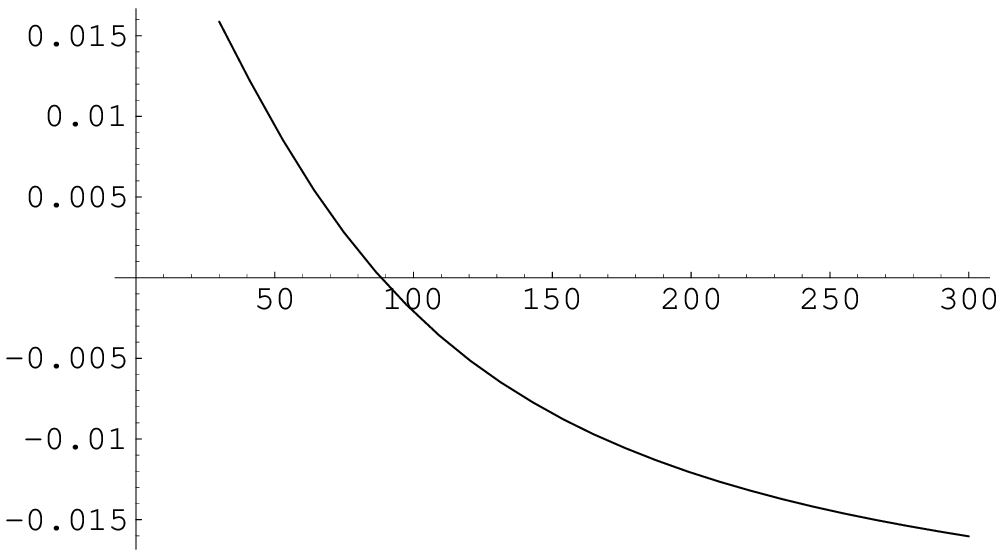}
 \hspace{0.0cm} $m_{LSP}\rightarrow$ ($GeV$)
\rotatebox{90}{\hspace{1.0cm} $h\rightarrow$}
\includegraphics[height=0.1\textheight]{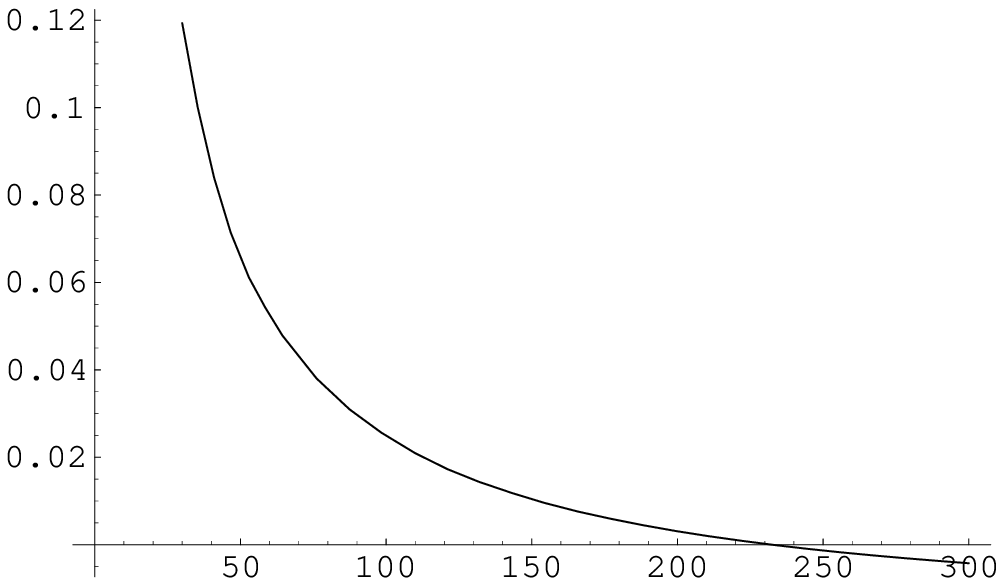}
 \hspace{0.0cm} $m_{LSP}\rightarrow$ ($GeV$)
\caption{ The the same as in Fig. \ref{ratio} for the modulation
amplitude $h$
 for the transition to the ground state. For the definitions see text.
\label{hgs} }
\end{center}
\end{figure}
\begin{figure}
\begin{center}
\rotatebox{90}{\hspace{1.0cm} $h_{exc}\rightarrow$}
\includegraphics[height=.10\textheight]{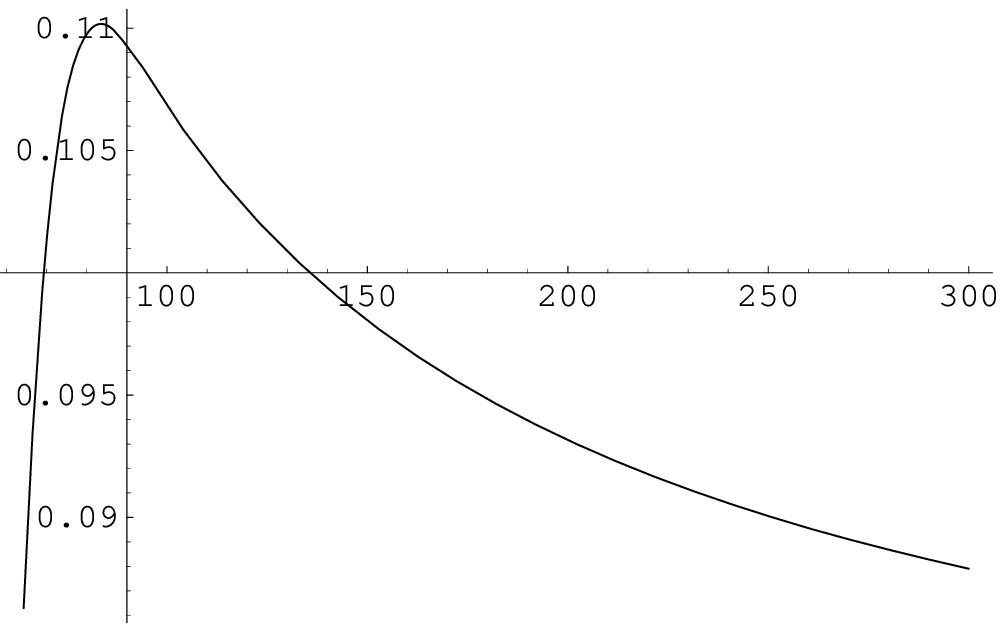}
 \hspace{0.0cm} $m_{LSP}\rightarrow$ ($GeV$)
\rotatebox{90}{\hspace{1.0cm} $h_{exc}\rightarrow$}
\includegraphics[height=.10\textheight]{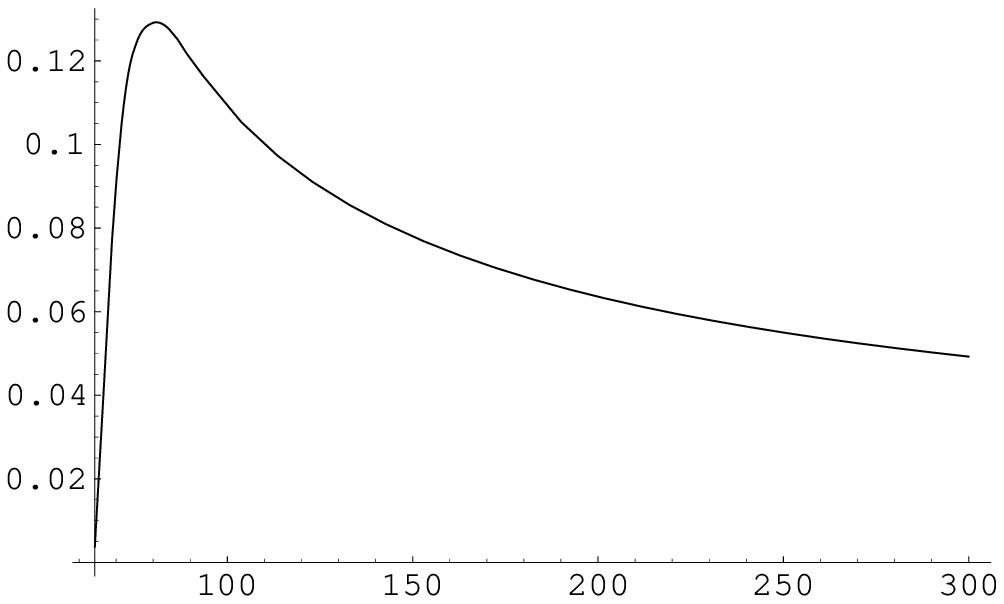}
\hspace{0.0cm} $m_{LSP}\rightarrow$ ($GeV$)
 \caption{ The the same
as in Fig. \ref{hgs} for the modulation amplitude $h_{exc}$ for
the transition to the excited state. \label{hexc} }
\end{center}
\end{figure}
 Finally from Figs \ref{hgs} and \ref{hexc} we notice that,
 especially  for $m_{\chi}\succeq100~GeV$, the modulation amplitude
 for transitions to the excited state is much bigger compared
to that of the elastic scattering. We hope that this additional
signal can perhaps be exploited by the experimentalists to
discriminate against background $\gamma$ rays, in a fashion
analogous to the ongoing measurements of nuclear recoils.
\par
Acknowledgments: The work of JDV  was supported in part by the
European Union under the contracts RTN No HPRN-CT-2000-00148 and
MRTN-CT-2004-503369. JDV and PhQ are indebted to Dr Dan
Strottman for his support and hospitality. All authors are
indebted to Professor Hiro Ejiri for his careful reading of the
manuscript and his invaluable comments related to the experimental
implications of this work.

\begin{table}[t]
\caption{ The static spin matrix elements for various nuclei. For
light nuclei the calculations are from Divari et al \cite{DIVA00}
. For $^{127}I$ the results are from Ressel and Dean \cite{Ress}
(*) and Homlund {\it et al} \cite {SUHONEN03}(**).
 For $^{207}Pb$ they were obtained previously \cite{ref1}, \cite{KVprd}.
}
\label{table.spin}
\begin{center}
\begin{tabular}{lrrrrrr}
 &   &  &  &  &   & \\
 & $^{19}$F & $^{29}$Si & $^{23}$Na  & $^{127}I^*$ & $ ^{127}I^{**}$ & $^{207}Pb^+$\\
\hline
    &   &  &  &  &    \\
$[\Omega_{0}(0)]^2$         & 2.610   & 0.207  & 0.477  & 3.293   &1.488 & 0.305\\
$[\Omega_{1}(0)]^2$         & 2.807   & 0.219  & 0.346  & 1.220   &1.513 & 0.231\\
$\Omega_{0}(0)\Omega_{1}(0)$& 2.707   &-0.213  & 0.406  &2.008    &1.501&-0.266\\
$\mu_{th} $& 2.91   &-0.50  & 2.22  &    &\\
$\mu_{exp}$& 2.62   &-0.56  & 2.22  &    &\\
$\frac{\mu_{th}(spin)}{ \mu_{exp}}$& 0.91   &0.99  & 0.57  &    &  &\\
\end{tabular}
\end{center}
\end{table}
\end{document}